\documentclass[prl,twocolumn,showpacs,floats]{revtex4}
\usepackage{graphicx}

\begin{document}

% The title goes here
%
\title{Identification of a New Form of Electron Coupling in Bi$_2$Sr$_2$CaCu$_2$O$_8$  Superconductor by Laser-Based Angle-Resolved Photoemission}
%
%
% The list of authors
%
\author{Wentao Zhang$^{1}$, Guodong Liu$^{1}$, Lin Zhao$^{1}$, Haiyun Liu$^{1}$, Jianqiao Meng$^{1}$, Xiaoli Dong$^{1}$, Wei Lu$^{1}$, J. S. Wen$^{2}$, Z. J. Xu$^{2}$, G. D. Gu$^{2}$, T. Sasagawa$^{3}$, Guiling Wang$^{4}$, Yong Zhu$^{5}$, Hongbo Zhang$^{4}$,Yong Zhou$^{4}$, Xiaoyang Wang$^{5}$, Zhongxian Zhao$^{1}$, Chuangtian Chen$^{5}$, Zuyan Xu$^{4}$ and X. J. Zhou $^{1,*}$}

\affiliation{
\\$^{1}$National Laboratory for Superconductivity, Beijing National Laboratory for Condensed Matter Physics, Institute of Physics, Chinese Academy of Sciences, Beijing 100080, China
\\$^{2}$Condensed Matter Physics and Materials Science Department, Brookhaven National Laboratory, Upton, New York 11973, USA
\\$^{3}$Materials and Structures Laboratory, Tokyo Institute of Technology, Yokohama Kanagawa, Japan
\\$^{4}$Laboratory for Optics, Beijing National Laboratory for Condensed Matter Physics,Institute of Physics, Chinese Academy of Sciences, Beijing 100080,  China
\\$^{5}$Technical Institute of Physics and Chemistry, Chinese Academy of Sciences, Beijing 100080, China}
%
%\date{Today}
%

\begin{abstract}

Laser-based angle-resolved photoemission measurements with
super-high resolution have been carried out on an optimally-doped
Bi$_2$Sr$_2$CaCu$_2$O$_8$ high temperature superconductor.  New high
energy features at $\sim$115 meV and $\sim$150 meV, besides the
prominent $\sim$70 meV one, are found to develop in the nodal
electron self-energy in the superconducting state. These high energy
features, which can not be attributed to electron coupling with
single phonon or magnetic resonance mode, point to the existence of
a new form of electron coupling in high temperature superconductors.

\end{abstract}

\pacs{74.72.Hs, 74.25.Jb, 79.60.-i, 71.38.-k}

\maketitle

The physical properties of materials are dictated by the microscopic
electron dynamics that relies on the many-body effects, i.e., the
interactions of electrons with other excitations, like phonons,
magnons and so on.  How to detect and disentangle these many-body
effects is critical to understand the macroscopic physical
properties and the superconductivity mechanism in high temperature
superconductors.  Angle-resolved photoemission spectroscopy (ARPES),
as a powerful tool in probing many-body effects\cite{ThreeReviews},
has revealed clear evidence of electron coupling with low energy
collective excitations (bosons) at an energy scale of
$\sim$70meV\cite{Bogdanovkink,Johnsonkink,Kaminskikink,Lanzarakink,Zhoukink,KordyukT}
in the nodal region and $\sim$40meV near the antinodal
region\cite{GromkoANKink,Kaminskikink,KimANKink,CukANKink} although
the nature of the bosonic modes remains under debate as to whether
it is phonon\cite{Lanzarakink,Zhoukink,CukANKink} or magnetic
resonance
mode\cite{Johnsonkink,Kaminskikink,KordyukT,EshrigT,GromkoANKink}.
Recently, another high energy feature is identified in dispersion at
300$\sim$400meV but its origin remains unclear as to whether this
can be attributed to a many-body effect\cite{Kink400meV}.

In this paper we report an identification of a new form of electron
coupling in high temperature superconductors by taking advantage of
super-high resolution vacuum ultra-violet (VUV) laser-based ARPES
technique\cite{LiuIOP}. New features at energy scales of
$\sim$115meV and $\sim$150meV are revealed in the electron
self-energy in Bi$_2$Sr$_2$CaCu$_2$O$_8$ (Bi2212) superconductor in
the superconducting state. These features can not be attributed to
electron coupling with single phonon mode or magnetic resonance
mode. They point to a possibility of electron coupling with some
high energy excitations in high temperature superconductors.

\begin{figure}[tbp]
\begin{center}
\includegraphics[width=1.00\columnwidth,angle=0]{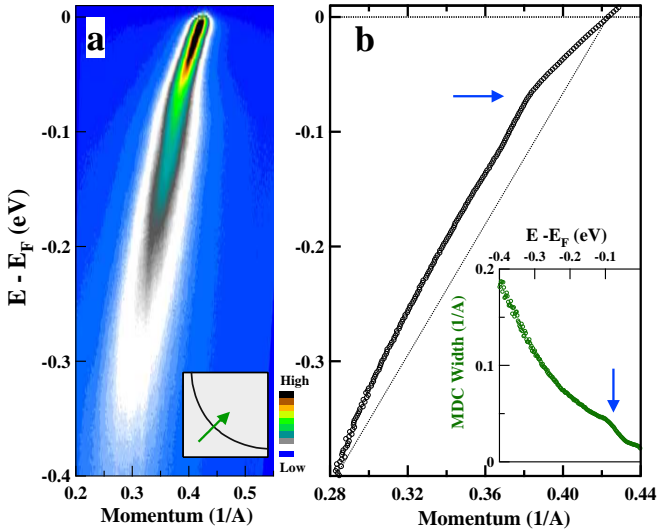}
\end{center}
\caption{Electron dynamics of optimally-doped Bi2212 (T$_c$=91K)
measured along the $\Gamma$(0,0)-Y($\pi$,$\pi$) nodal direction at
17K. (a). Raw image showing photoelectron intensity (represented by
false color) as a function of energy and momentum. The inset shows
the location of the  momentum cut in the Brillouin zone; (b). Nodal
dispersion extracted from Fig. 1a by fitting MDCs. The dotted line
connecting the two energy positions in the dispersion at the Fermi
energy and -0.4 eV  is an empirical bare band for extracting the
effective real part of self-energy in Fig. 2b. The inset shows the
corresponding MDC width (Full-Width-at-Half-Maximum, FWHM). }
\end{figure}

The angle-resolved photoemission measurements have been carried out
on our newly-developed VUV laser-based angle-resolved photoemission
system\cite{LiuIOP}. The photon energy of the laser is 6.994 eV with
a bandwidth of 0.26 meV. The energy resolution of the electron
energy analyzer (Scienta R4000) is set at 0.5 meV, giving rise to an
overall energy resolution of 0.56 meV which is significantly
improved from 10$\sim$15 meV  from regular synchrotron radiation
systems\cite{Bogdanovkink,Johnsonkink,Kaminskikink,Lanzarakink,
Zhoukink,KordyukT}.  The angular resolution is $\sim$0.3$^\circ$,
corresponding to a momentum resolution  $\sim$0.004 $\AA$$^{-1}$ at
the photon energy of 6.994 eV, more than twice improved from 0.009
$\AA$$^{-1}$ at a regular photon energy of 21.2 eV for the same
angular resolution. The photon flux is adjusted between 10$^{13}$
and 10$^{14}$ photons/second. The optimally-doped Bi2212 single
crystals with a superconducting transition temperature T$_c$=91K
were cleaved {\it in situ} in vacuum with a base pressure better
than 5$\times$10$^{-11}$ Torr.

Fig. 1a shows the raw data of photoelectron intensity as a function
of energy and momentum for an optimally-doped (Bi2212)
superconductor (T$_c$=91K) measured along the
$\Gamma$(0,0)-Y($\pi$,$\pi$) nodal direction at a temperature of
17K.  By fitting momentum distribution curves (MDCs), the dispersion
(Fig. 1b) and MDC width (inset of Fig. 1b) are quantitatively
extracted from Fig. 1a. One can see an obvious kink in dispersion
near 70 meV (Fig. 1a and b) and a drop in the MDC width(inset of
Fig. 1b), similar to those reported
before\cite{Bogdanovkink,Johnsonkink,Kaminskikink,Lanzarakink,Zhoukink,KordyukT}
but with much improved clarity. It is generally agreed that this 70
meV feature originates from a coupling of electrons with a
collective boson mode. When coming to the nature of the boson mode,
it remains under debate whether it is
phonon\cite{Lanzarakink,Zhoukink} or magnetic resonance
mode\cite{Johnsonkink,Kaminskikink,KordyukT}.

The real part of the electron self-energy can be extracted from the
dispersion given that the  bare band dispersion is known which can
be determined in a number of ways but still without
consensus\cite{KordyukBBand,ZhouFS}. To identify fine features in
the electron self-energy and study their relative change with
temperature, it is reasonable to assume a featureless bare band for
the nodal dispersion within a small energy window near the Fermi
energy. In this case, the fine features manifest themselves either
as peaks or curvature changes in the ``effective
self-energy"\cite{ZhouFS}.  As shown in Fig. 1b,  we choose here a
straight line connecting two energy positions in the dispersion at
the Fermi energy and -0.4eV as the empirical bare band. The
resultant effective real part of electron self-energy is shown in
Fig. 2b.  Also shown in Fig. 2 are dispersions (Fig. 2a)  along
several other cuts in the Brillouin zone (inset of Fig. 2a) and the
corresponding effective electron self-energy (Fig. 2b).

\begin{figure}[tbp]
\begin{center}
\includegraphics[width=1.00\columnwidth,angle=0]{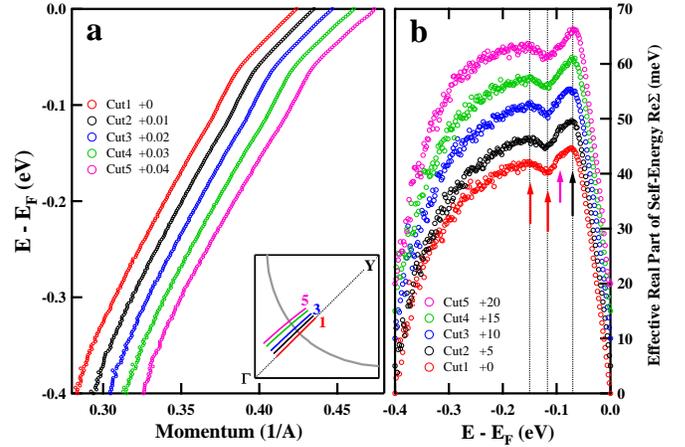}
\end{center}
\caption{Momentum dependence of dispersions(a) and corresponding
effective real part of electron self-energy (b). The inset of Fig.
2a shows the location of momentum cuts in the Brillouin zone. The
effective self-energy in Fig. 2b is obtained by taking the straight
line connecting two positions in dispersion at the Fermi energy and
-0.4eV as a bare band. The arrows in Fig. 2b mark fine features at
$\sim$70 meV, 115 meV, and 150 meV, and a possible feature near 94
meV. For clarity, curves in Fig. 2a is offset along the momentum
axis while curves in Fig. 2b is offset along the vertical axis; the
offset values are given in the legends. }
\end{figure}

With much improved precision of data, one can identify clearly
several features in the electron self-energy, as shown in Fig. 2b.
The most pronounced feature is the peak at $\sim$ 70 meV that gives
rise to the kink in dispersion seen here and
before\cite{Bogdanovkink,Johnsonkink,Kaminskikink,Lanzarakink,Zhoukink,KordyukT}.
In addition, at higher energies, two new features can be identified
clearly as a valley at $\sim$115meV and a cusp at $\sim$150 meV.
Signature of a fine feature near 94 meV is also visible,
particularly for the two cuts close to nodal region (cuts 1 and 2).
Between the Fermi level and 70 meV, we have also observed hints of
possible low-energy features which are however very subtle and need
further measurements to pin them down. We note that the bare band
selection has little effect on the identification of these fine
features and their energy position as we have checked by trying
other straight lines as empirical bare bands. Particularly, the two
new features at 115 meV and 150 meV, together with the 70 meV peak,
are robust (as also shown in Fig. 4 below) and persist in a rather
large momentum space near the nodal region.

The nodal electron dynamics undergoes a dramatic evolution with
temperature and superconducting transition, as indicated by the
temperature dependence of the nodal dispersion (Fig. 3a) and
scattering rate (Fig. 3b).  In Fig. 3a, a quantitative momentum
variation with temperature at four typical energy positions (the
Fermi level, -0.07 eV, -0.2 eV and -0.3 eV) is plotted in the
upper-left inset and some representative MDCs for these four
energies at 17K and 128K are plotted in the bottom-right inset.
Over the temperature range of the measurement, the dispersion change
with temperature spreads over an energy range of  0$\sim$300meV
within which the dispersion renormalization gets stronger with
decreasing temperature.

An unexpected finding is the Fermi momentum shift with temperature
(Fig. 3a and top-left inset) which increases with increasing
temperature in the superconducting state below T$_c$=91K and then
becomes nearly flat above T$_c$. The magnitude of the shift is
small, on the order of 0.003 $\AA$$^{-1}$, and the change is
monotonic. We first checked whether this could be caused by sample
orientation change during heating/cooling process and feel that it
is unlikely because this usually would cause a shift of overall
dispersion. As shown in Fig 3a and the top-left inset, the
dispersion above 300 meV shows little change with temperature. In
fact, the MDCs themselves at 300 meV overlap with each other at 17K
and 128K almost perfectly, as shown in the bottom-right inset of
Fig. 3a. Another possibility we have checked is whether it can be
caused by the thermal expansion or contraction of the sample during
temperature change. This can also be excluded because with the
lattice expansion with increasing temperature, one would expect a
shrink of the Brillouin zone that causes the nodal Fermi momentum
move to smaller value; this expected trend is just opposite to our
experimental observation. Because the Fermi momentum shift with
temperature is quite unusual and has not been reported before, we
have repeated the measurement and  reproduced the similar
observation. Therefore, to the best of our effort, we think this is
likely an intrinsic effect and its first observation is a result of
much improved instrumental precision. Further work needs to be done
to pin down this effect and understand the underlying physical
origin as it suggests either a chemical potential shift or Fermi
surface topology change upon entering the superconducting state.

\begin{figure}[tbp]
\begin{center}
\includegraphics[width=1.00\columnwidth,angle=0]{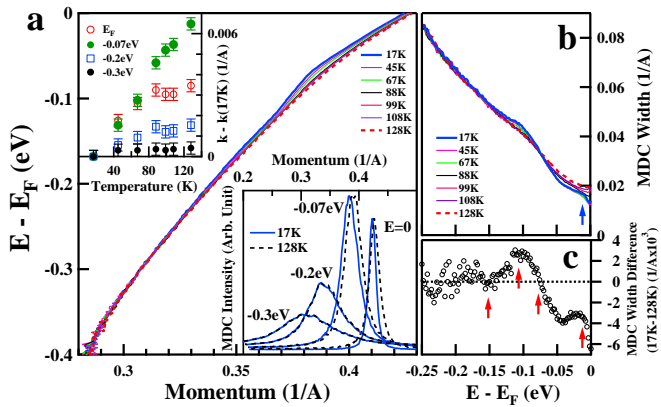}
\end{center}
\caption{Temperature dependence of the nodal MDC dispersion (a) and
MDC width (b). The top-left inset of Fig. 3a plots the momentum
value as a function of temperature at four typical energies: E$_F$
(red empty circle), -0.07 eV (green solid circle), -0.2 eV (blue
empty square) and -0.3 eV (black solid circle) obtained by averaging
over $\pm$10 meV energy range . The bottom-right inset of Fig. 3a
shows MDCs at these four energies measured at 17K (solid line) and
128K (dashed line). (c). The difference of the MDC width between two
temperatures at 17K and 128K. Arrows in Fig. 3b and c mark possible
features showing up in the scattering rate.}
\end{figure}

The scattering rate shows a strong variation with temperature at low
energy range within 0$\sim$-0.2 eV, as shown in Fig. 3b.
Interestingly, the temperature dependence is not monotonic but
depends on the binding energy.  Between the Fermi level and -0.07
eV, the scattering rate decreases with decreasing temperature,
while it increases with decreasing temperature between -0.07 and
-0.15 eV. This gives rise to an ``overshoot" region extending to
$\sim$-0.1 eV at low temperature.  The change of the scattering rate
with temperature can also be clearly seen in the difference between
the normal and superconducting states, as plotted in Fig. 3c which
depicts the difference between 17K and 128K data.  Here the
difference between -0.07 eV and -0.15 eV is positive while it
becomes negative between E$_F$ and -0.07 eV.

The temperature dependence of the effective real part of self-energy
(Fig. 4) indicates that the new high energy features at 115 meV and
150 meV are developed in the superconducting state.  Fig. 4a shows
the effective real part of self-energy at various temperatures
obtained from the dispersions (Fig. 3a) by selecting a common
empirical bare band, i.e., a straight line connecting the Fermi
energy and -0.4 eV in the dispersion at 128K.  Fig. 4b shows the net
temperature change of electron self-energy with respect to the
normal state data at 128K,  thus avoiding any ambiguity from bare
band selection.  It is clear in both cases that dramatic change of
electron self-energy occurs in the superconducting state, with a
sharpening of the $\sim$70 meV feature, together with the emerging
and growing of the $\sim$115meV and $\sim$150meV features.   These
observations are consistent with the scattering rate data where one
can see the emergence of similar characteristic energy scales at
$\sim$150 meV, $\sim$115 meV and $\sim$70 meV in the superconducting
state (Fig. 3c).

\begin{figure}[tbp]
\begin{center}
\includegraphics[width=1.00\columnwidth,angle=0]{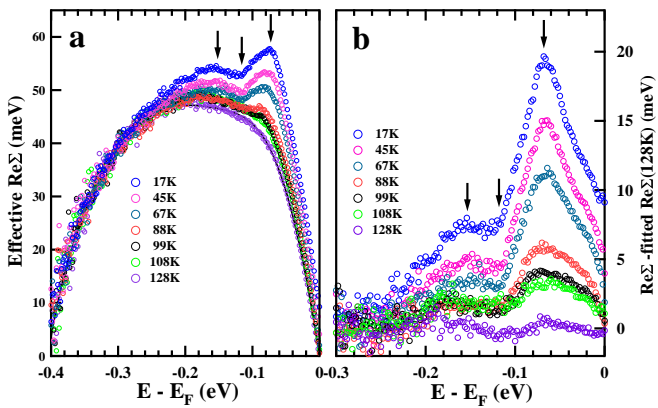}
\end{center}
\caption{(a).Temperature dependence of the effective real part of
electron self-energy extracted from dispersions in Fig. 3a by taking
a straight line as a bare band connecting two points at E$_F$ and
-0.4eV on the dispersion at 128K.  (b). Temperature dependence of
the difference between the measured self-energy in Fig. 4a and the
fitted one for 128K (solid black line in Fig. 4a). The 128K
self-energy is fitted by polynomials and used to reduce statistical
errors. }
\end{figure}

The identification of high energy features at 115 meV and 150 meV
points to a new form of electron coupling in high temperature
superconductors. These are qualitatively different from the $\sim$70
meV feature which is attributed to coupling of electrons with some
collective modes and such modes with comparable energy scales are
available in high temperature superconductors either as
phonon\cite{Lanzarakink,Zhoukink} or magnetic resonance
mode\cite{Johnsonkink,Kaminskikink,KordyukT}. Because the energy
scale of these two new features is higher than the maximum energy of
phonons ($\sim$90 meV)\cite{PhononMax} and magnetic resonance mode
(42 meV in optimally-doped Bi2212)\cite{HeResonance}, they can not
be due to electron coupling with a single phonon or magnetic
resonance mode.  Although the effect of the electron-boson coupling
for a low-energy mode can extend to high energy in the electron
self-energy, it will not generate any new features with clear
curvature change, as evidenced from simulations of electron-phonon
coupling using both Debye and Einstein models and confirmed in
canonical electron-phonon coupling systems\cite{BeRef}.

There are a couple of possibilities that may give rise to high
energy features in the electron coupling. The first is the mode
energy shift due to the opening of superconducting gap: the original
mode position in the normal state is expected to be shifted upward
by an amount on the order of the superconducting gap upon entering
the superconducting state\cite{Sandvik}. In the optimally-doped
Bi2212, with the maximum d-wave superconducting gap at $\sim$35meV,
one might expect that the original $\sim$70meV mode be shifted to a
higher energy around 105 meV. However, this scenario is difficult to
explain the existence of another 150 meV feature, the feature at 110
meV being a valley instead of a peak and the remaining strong
$\sim$70meV feature. Another possibility is the electron coupling
with multiple phonons. Usually this effect is expected to be much
weaker\cite{Engelsberg} although in principle the possibility can
not be totally excluded. More theoretical work are needed to verify
whether such multiphonon process can produce clear features at high
energy and whether such an effect is enhanced at low temperature.
The third possibility is electron coupling with excitations that are
already present at such high energy scales. The emergence and
evolution of the 115meV and 150meV features in the electron
self-energy is probably due to the redistribution of the underlying
spectral function of the high energy excitations with temperature
and superconducting transition.  One candidate of such high energy
excitations in high temperature superconductors seems to be
naturally related to the spin fluctuation observed by neutron
scattering, which covers a large energy range up to 200 meV and
exhibits strong temperature dependence\cite{NeutronIS}. Signature of
such high energy coupling is also proposed from optical
measurements\cite{Carbotte}. Further theoretical work to investigate
the effect of such high energy spin excitations on electron dynamics
will help in clarifying such a scenario.

In conclusion, by performing high precision ARPES measurements on
Bi2212, we have revealed new features at 115 meV and 150 meV in the
electron self-energy developed in the superconducting state. They
can not be attributed to electron coupling with either single phonon
or magnetic resonance mode but point to the existence of a new form
of electron coupling in high temperature superconductors. We hope
this observation will stimulate further theoretical work to
understand their origin and their role in determining anomalous
physical properties of high temperature superconductors.

We acknowledge helpful discussions with J. R. Shi, T. Xiang, Z. Y.
Weng and S. Kivelson. This work is supported by the NSFC, the MOST
of China (973 project No: 2006CB601002, 2006CB921302), and CAS
(Projects ITSNEM and 100-Talent).  The work at BNL is supported by
the DOE under contract No. DE-AC02-98CH10886.

$^{*}$Corresponding author (XJZhou@aphy.iphy.ac.cn)

\end{document}